\documentclass[10pt,conference]{IEEEtran}

\IEEEoverridecommandlockouts

\usepackage{cite}
\usepackage{amsmath,amssymb,amsfonts}
\usepackage{algorithm}
\usepackage{algorithmic}
\usepackage{graphicx}
\usepackage{textcomp}
\usepackage{xcolor}
\usepackage{listings}
\usepackage{amsmath}
\usepackage[normalem]{ulem}
\DeclareMathOperator*{\argmin}{arg\,min}

\usepackage[utf8]{inputenc}
\usepackage[english]{babel}
\usepackage[T1]{fontenc}
\usepackage{amsmath}
\usepackage{amsthm}
\usepackage{amsfonts}
\usepackage{hyperref}
\usepackage{graphicx}
\usepackage{braket}
\usepackage{enumerate}
\usepackage{multirow}
\usepackage{tikz}
\usepackage{enumitem}
\usepackage{soul}

\usepackage{array}
\newcolumntype{L}[1]{>{\raggedright\let\newline\\\arraybackslash\hspace{0pt}}m{#1}}
\newcolumntype{C}[1]{>{\centering\let\newline\\\arraybackslash\hspace{0pt}}m{#1}}
\newcolumntype{R}[1]{>{\raggedleft\let\newline\\\arraybackslash\hspace{0pt}}m{#1}}
\usepackage{makecell}
\usepackage{colortbl}

\usepackage[dvipsnames,svgnames,x11names]{xcolor}

\colorlet{RED}{red}

\usetikzlibrary{quantikz2}
\def\BibTeX{{\rm B\kern-.05em{\sc i\kern-.025em b}\kern-.08em
    T\kern-.1667em\lower.7ex\hbox{E}\kern-.125emX}}

\begin{document}

\title{Noise-aware selection of circuit cutting strategies \\under hardware noise non-uniformity}

\author{Debarthi Pal$^{1}$, Ritajit Majumdar$^{2}$,\\ Padmanabha Venkatagiri Seshadri$^{3}$, Anupama Ray$^{2}$,  Yogesh Simmhan$^{1}$\\~\\

$^1$Indian Institute of Science, Bangalore\\
$^2$\emph{IBM Quantum}, IBM India Research Lab\\
$^3$IBM Research, India\\

debarthipal@iisc.ac.in, majumdar.ritajit@ibm.com,\\ seshapad@in.ibm.com, anupamar@in.ibm.com,  simmhan@iisc.ac.in}

\maketitle

\pagestyle{plain}
\thispagestyle{plain}

\begin{abstract}

Noise in contemporary quantum hardware is highly non-uniform across qubits and couplers, giving rise to localized low-noise ``islands'' within otherwise noisy device topologies. As quantum workloads scale, executions are increasingly forced to traverse high-noise regions, degrading algorithmic fidelity. Circuit cutting provides a route to circumvent such regions by decomposing large circuits into smaller subcircuits, but its practicality is limited by exponential sampling overhead and the lack of systematic guidance on how cut strategies should align with heterogeneous hardware noise. In this work, we present a hardware-noise-aware circuit cutting framework that explicitly exploits the spatial non-uniformity of noise in quantum devices. Rather than proposing a new cut-finding algorithm, we formalize the problem of \emph{device-constraint selection} under realistic hardware noise and show that this choice critically determines both execution overhead and effective noise. Using a unified gate- and wire-cutting formulation, we demonstrate that small, hardware-informed relaxations in the device constraint yield exponential reductions in execution overhead while preserving alignment with low-noise hardware regions. Across representative workloads, our method achieves an average reduction in the number of circuit executions ranging from $5$--$54\times$ for 20-qubit circuits, and enables tractable circuit cutting for 50-qubit circuits and application-level benchmarks where conventional strategies incur prohibitive overhead. These results establish noise-aware device-constraint selection as a necessary ingredient for making circuit cutting resource-efficient and practically deployable on contemporary quantum hardware.

\end{abstract}


\IEEEpeerreviewmaketitle

\section{Introduction}
\label{sec:intro}
Quantum computing is tending towards utility-scale applications, which has increased the size of the quantum circuit workloads in terms of the number of qubits and circuit depth. These circuits can now be executed on quantum hardware with a large number of qubits (e.g., 133-qubit IBM Torino). However, noise is an artifact of quantum hardware and affects the performance of quantum algorithms. 
Importantly, noise characteristics are not uniform across the hardware topology: qubits and couplers exhibit varying fidelities, giving rise to localized regions of comparatively low noise. Before the utility-era, quantum circuit sizes were relatively smaller than the qubit topology of the hardware and could leverage the islands of good performance (low noise sub-topologies of the qubits) directly. In contrast, as circuit sizes approach the scale of available hardware, such favorable regions may no longer accommodate entire workloads, forcing execution to traverse noisier parts of the device. This motivates the need for methods that can \textit{exploit low-noise regions of the hardware even when the target circuit exceeds their size}. In this work, we explore the use of circuit cutting, in conjunction with hardware-awareness, as a means to avoid high-noise regions while executing large quantum circuits.

Circuit cutting partitions a quantum circuit into multiple smaller subcircuits whose outcomes can be recombined to either obtain probability distributions~\cite{tang2021cutqc} or expectation values~\cite{peng2020simulating}. Existing approaches include wire cutting, where circuits are partitioned along qubit wires~\cite{peng2020simulating, tang2021cutqc, brenner2025optimal, majumdar2022error}, and gate cutting, where multi-qubit gates are decomposed into samples of single-qubit operations~\cite{mitarai2021constructing, mitarai2021overhead, schmitt2025cutting}. These approaches can be unified under a Quasi-Probability Distribution (QPD) framework, where the sampling overhead $\gamma$ grows exponentially with the number of cuts (e.g., $\gamma = 9^{k}$ for $k$ gate cuts \cite{mitarai2021constructing} and $16^{k}$ for $k$ wire cuts~\cite{peng2020simulating}), making circuit cutting practical only when the number of cuts is carefully controlled.

Several approaches have been studied for this \cite{tang2021cutqc, basu2022qer, basu2024fragqc} which are briefly elucidated in Section~\ref{sec:related}. However, most of these consider only wire cutting, making them restrictive for certain circuits (see Table~\ref{tab:cutting_comparison}), or assume that the hardware noise profile is uniform, or leave key parameters, such as number of qubits in each subcircuit, to user choice. An automatic heuristic cut-finder capable of handling \textit{both wire and gate cuts} was introduced in~\cite{qiskit-addon-cutting}. This method accepts a device constraint, i.e., the maximum allowed subcircuit size, and produces a partitioning involving both gate and wire cuts that minimizes sampling overhead while satisfying the constraint.

\begin{figure}[t]
    \centering
    \includegraphics[width=1\columnwidth]{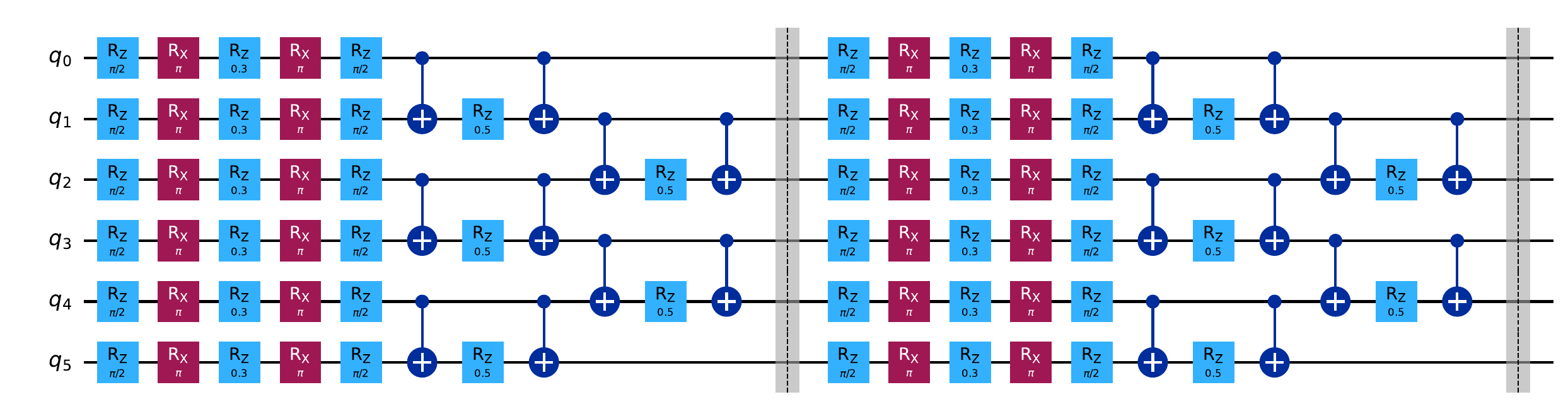}
    \caption{A 6-qubit 1D Ising model having nearest neighbour interactions with 2 trotter steps. The circuit execution overheads to partition this circuit for different device constraints are shown in Table~\ref{tab:cut_stats}.}
    \label{fig:example_ising}
\end{figure}

Crucially, the effectiveness of such automatic cut-finding algorithms depends strongly on the choice of the \emph{device constraint}. Selecting an overly conservative constraint can lead to unnecessary cuts and exponential growth in execution cost, whereas modest relaxations can yield disproportionate gains. Fig.~\ref{fig:example_ising} illustrates this for a 6-qubit 1D Ising circuit with two Trotter steps, while Table~\ref{tab:cut_stats} shows that increasing the device constraint from 3 (i.e., equal partitioning, the most natural `naive' choice) to 4 reduces the number of required circuit executions from 6561 to 1296; whereas reducing it to 2 increases the number of circuit executions to 43M. These observations indicate that device-constraint selection is a critical but underexplored optimization lever for practical circuit cutting.

\begin{table}[t]
\centering
\caption{Impact of device constraint on the number of cuts and the number of circuit executions for the circuit in Fig.~\ref{fig:example_ising}}
\begin{tabular}{c|c|c|r}
\hline
\textbf{Device constraint} & \multicolumn{2}{c|}{\textbf{Types of Cuts}} & \textbf{\# Circuit Executions} \\
\cline{2-3}
 & \textbf{Gate Cut} & \textbf{Wire Cut} &  \\
\hline\hline
2 & 8 & 0 & 43,046,721 \\\hline
3 & 4 & 0 & 6,561 \\\hline
4 & 2 & 1 & 1,296 \\
\hline
\end{tabular}
\label{tab:cut_stats}
\end{table}

However, a key gap remains in \textit{how to systematically select device constraints under realistic, non-uniform hardware noise so that resulting subcircuits align with low-noise regions of the hardware while controlling sampling overhead}. Existing approaches leave this decision largely to user intuition or rely on uniform-noise assumptions that do not hold in practice. In this work, we address this limitation by introducing a hardware-noise-aware methodology for selecting circuit cutting strategies. We leverage spatial noise heterogeneity to guide device-constraint selection in automatic cut finding, producing subcircuits that better conform to low-noise sub-layouts of the hardware topology while limiting execution overhead. 

Specifically, we propose \emph{Hardware-Inspired Cutting (HIC)} where we: 
\begin{enumerate}[leftmargin=*]
    \item Demonstrate that circuit cutting performance under realistic noise is highly sensitive to device constraint selection, with small changes leading to exponential differences in sampling overhead;
    \item Propose a noise-aware, quantum-centric algorithm for device-constraint selection that aligns cut-induced subcircuits with low-noise regions of the hardware topology; and
    \item Evaluate HIC across structured, random and application-level workloads using real device noise data, showing that it enables substantially lower execution overhead than noise-agnostic cutting strategies while preserving result quality whenever feasible.    
\end{enumerate}

The rest of the paper is organized as follows. Section~\ref{sec:our-proposed-method-of-cutting} introduces HIC, including hardware coupling map puncturing and device-constraint selection based on connected components; 
Section~\ref{sec:result} presents experimental results on 20- and 50-qubit circuits, along with application-level benchmarks from the Benchpress suite~\cite{nation2025benchmarking}; Section~\ref{sec:related} contrasts our work against existing literature, and Section~\ref{sec:cutqc_vs_hic} compares the pre-processing and post-processing time, and the execution overhead of HIC against existing cut finding approaches. Section~\ref{sec:conclusion} concludes with a discussion and future directions.

\section{Hardware-Inspired Cutting (HIC) Methodology}
\label{sec:our-proposed-method-of-cutting}

This section presents the Hardware-Inspired Cutting (HIC) methodology for selecting circuit cutting strategies under realistic hardware noise non-uniformity. 
Importantly, HIC does not directly search for optimal cut locations. Rather, it systematically selects hardware-informed \emph{device constraints} that guide existing automatic cut finders toward noise-resilient and resource-efficient cut strategies.
HIC is orthogonal to prior work on cut-location optimization, and instead targets the previously unaddressed problem of systematic device-constraint selection under non-uniform hardware noise.

\begin{figure}[t]
    \centering
    \includegraphics[width=0.9\columnwidth]{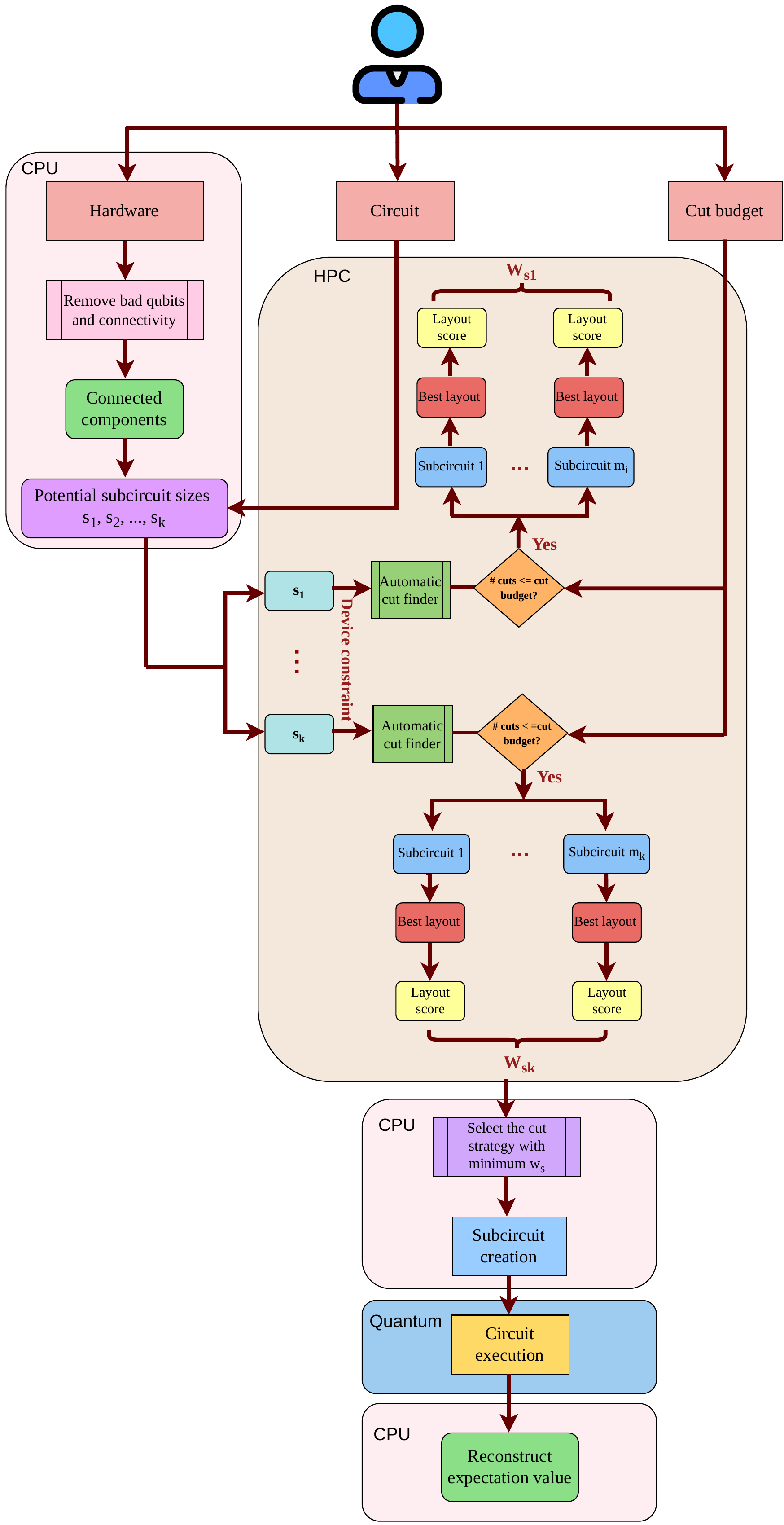}
    \caption{Workflow for HIC: the process involves pruning hardware based on qubit and connectivity noise profile, identifying connected components, and parallelizing the automatic cut finder across various potential subcircuit sizes to select the strategy with the minimum weighted layout score $W_s$. This figure explicitly demonstrates a quantum-centric algorithm to circuit cutting.}
    \label{fig:hic_flowchart}
\end{figure}

When a quantum circuit is mapped to the underlying hardware topology, called \emph{transpilation}, the goal of the transpiler is to minimize the number of SWAP gates as well as select the qubits and connectivities with lower noise~\cite{li2019tackling, sivarajah2020t, lightsabre}. If a circuit is placed on a particular layout $l$, then a function can be assigned which maps the noise-profile of layout $l$ to a real number, called \emph{layout score}~\cite{nation2023suppressing}. In all our experiments in this paper, we use the convention of~\cite{nation2023suppressing} where \textit{lower score implies better layout quality}. However, other scoring methods~\cite{majumdar2025sparse, srivastava2025lightweight} can have different conventions. HIC builds on this insight by explicitly incorporating hardware noise non-uniformity into device-constraint selection.

Current quantum devices exhibit significant spatial variation in qubit and connectivity quality. While transpilers attempt to avoid noisy components, large circuits often force execution onto such regions due to size constraints. 
HIC addresses this limitation through a three-stage workflow illustrated in Fig.~\ref{fig:hic_flowchart}: (i) \textit{Puncturing} the hardware coupling map to remove noise outliers, (ii) \textit{Enumerating} feasible device constraints from the resulting connected components, and (iii) \textit{Selecting} the cut strategy that minimizes a weighted layout score. This workflow ensures that generated subcircuits align with low-noise hardware regions while controlling sampling overhead.

\subsection{Constructing the Punctured Coupling Map}
\begin{figure}[t]
    \centering
    \includegraphics[width=0.9\columnwidth]{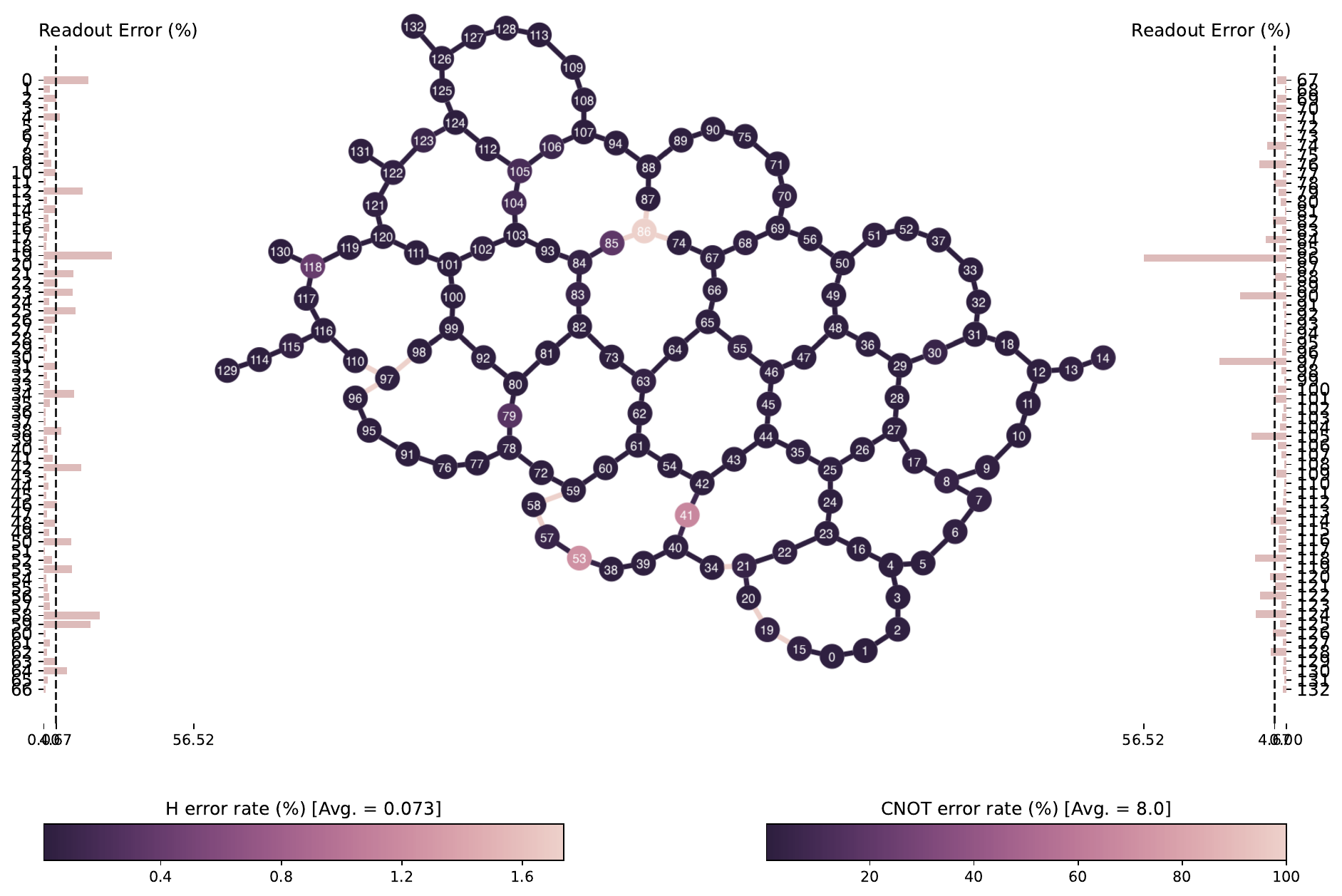}
    \caption{Noise profile of a 133 qubit IBM Quantum Heron device showing the non-uniformity of noise for both qubits and 2-qubit connectivity}
    \label{fig:noise_profile}
\end{figure}

A coupling map is a connected graph $G = (V,E)$, where the vertices represent the physical qubits and the edges represent the 2-qubit connectivities. Current quantum devices are noisy -- thus each of the qubits and connectivities have some error rate associated with them. In Fig.~\ref{fig:noise_profile} we show the noise profile of a 133-qubit IBM Quantum Heron device.

\begin{figure}[t]
    \centering
    \includegraphics[width=0.9\columnwidth]{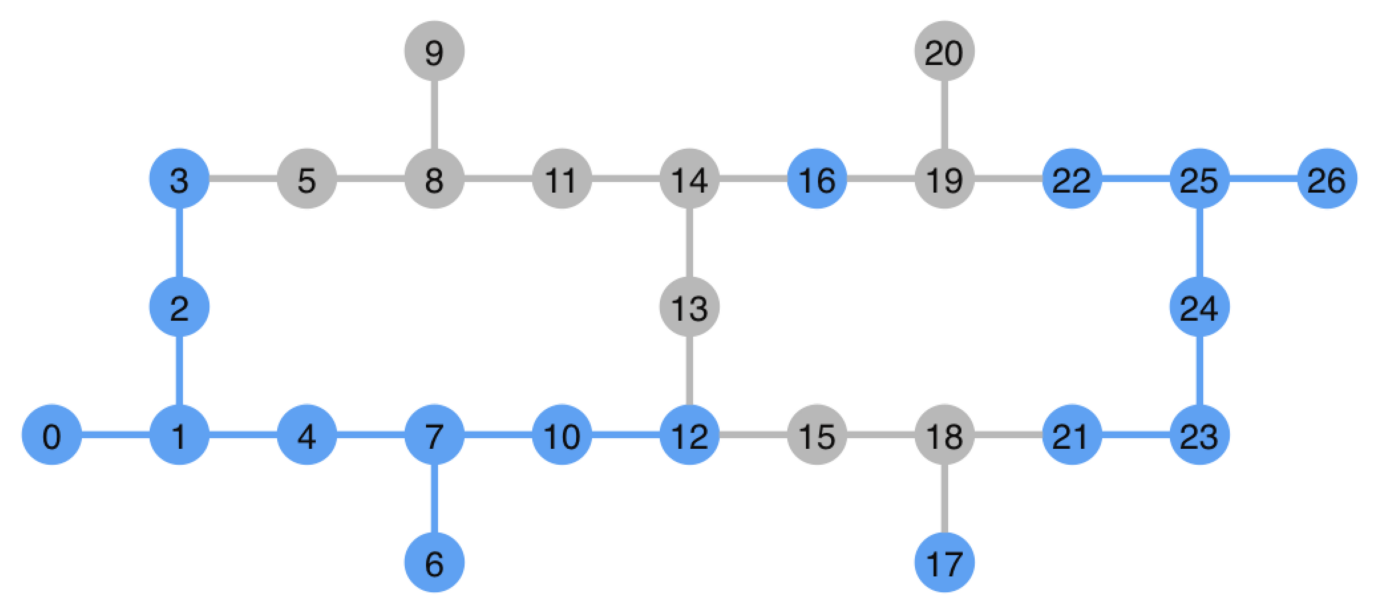}
    \caption{Punctured coupling map of a 27-qubit IBM Quantum device obtained after removing outlier qubits and connectivities from the hardware coupling map where $z = 0.05$. The blue and gray qubits (edges) respectively denote the ones retained and removed after puncturing. Dangling edges are naturally always removed irrespective of their error rate.}
    \label{fig:punctured_coupling_map}
\end{figure}

Outlier qubits and connectivities can be selected in different methods. One of the possible methods, which we use in this paper, is \emph{Z-score}, where Z-scores are computed independently for a single hardware calibration snapshot used in each experiment. The Z-score is given by
%
    $Z = \frac{x - \mu}{\sigma}$,
%
where $x$ is the error rate of a qubit (or 2-qubit connectivity), and $\mu$ and $\sigma$ are, respectively, the mean and standard deviation of the error rates over all the qubits (or 2-qubit connectivities). We pre-determine a value of $Z = z > 0$, and eliminate all $x$ for which $Z > z$. Note that this removes outliers which are at least $z\cdot\sigma$ away from the mean $\mu$. We repeat this separately for each qubit and 2-qubit connectivity. This ensures that the remaining qubits and connectivities have a noise profile closer to the mean, i.e., filters noise outliers while preserving the bulk noise characteristics of the device. 
This creates a \emph{punctured coupling map}, which may become disjoint -- creating multiple connected components. Algorithm~\ref{alg:punctured_coupling_map} lists the steps to obtain the punctured coupling map of a given hardware device, and Fig.~\ref{fig:punctured_coupling_map}  shows an example for a 27-qubit IBM Quantum device, where $z = 0.05$.

\begin{algorithm}[t]
\caption{Construct Punctured Coupling Map}
\label{alg:punctured_coupling_map}
\footnotesize
\begin{algorithmic}[1]
    \REQUIRE (i) hardware $h$; (ii) pre-determined Z-scores $z_V$ and $z_E$ for qubits and 2-qubit connectivities
    \ENSURE punctured coupling map $P_h$ of $h$

    \STATE $C_h = (V, E) \leftarrow$ coupling map of $h$ where $V$ and $E$ denote the sets of qubits and 2-qubit connectivities respectively
    \STATE $N_v \leftarrow$ noise profile of each qubit $v \in V$
    \STATE $N_e \leftarrow$ noise profile of each 2-qubit connectivity $e \in E$
    \STATE $\mu_V, \sigma_V \leftarrow$ mean and standard deviation of the noise profile of all qubits in $V$
    \STATE $\mu_E, \sigma_E \leftarrow$ mean and standard deviation of the noise profile of all 2-qubit connectivities in $E$
    \STATE bad\_qubits, bad\_edges $\leftarrow$ [], []
    \FORALL{$v \in V$}
        \IF{$\frac{N_v - \mu_V}{\sigma_V} > z_V$}
            \STATE add $v$ to bad\_qubits
        \ENDIF
    \ENDFOR
    \FORALL{$e \in E$}
        \IF{$\frac{N_e - \mu_E}{\sigma_E} > z_E$}
            \STATE add $e$ to bad\_edges
        \ENDIF
    \ENDFOR
    \STATE $P_h \leftarrow$ punctured coupling map obtained by removing the qubits in bad\_qubits and 2-qubit connectivities in bad\_edges from $C_h$ and any dangling edge.
    \RETURN $P_h$
\end{algorithmic}

\end{algorithm}

The noise profile of the qubits and the 2-qubit connectivities can be obtained from the calibration data of the hardware, or by executing benchmark experiments to obtain the updated noise information. 
Each connected component of the punctured coupling map can be interpreted as an \emph{``island of comparatively low noise''}. For large circuits, the connected components of the punctured coupling map will not be large enough to accommodate it. 
Consequently, the size of a connected component serves as a natural \textit{upper bound} on the number of qubits in a subcircuit, and is used as the value of the \emph{device constraint} supplied to the automatic cut finder~\cite{qiskit-addon-cutting}.

\subsection{Selection of Cut-strategy from Punctured Coupling Map}
\label{sec:cut-strategy}

In Algorithm~\ref{alg:cut_strategy} we propose a method to find the cut strategy for a given circuit. From the above punctured coupling map, this algorithm filters all the connected components that can support one or more of the subcircuits. We call the number of qubits in each of the filtered connected components as the \emph{potential subcircuit size} since these serve as the upper bound for the size of a subcircuit, post cutting. Each such size is treated as a candidate \emph{device constraint} that can be supplied to the \textit{automatic cut finder}~\cite{qiskit-addon-cutting}. For each subcircuit size, the automatic cut finder searches for an optimal cutting strategy. We retain those cutting strategies where the number of cuts is within a pre-defined cut budget $k_{max}$. Since the evaluation of each candidate device constraint is independent of the others, this can be executed in parallel, enabling efficient exploration of the constraint space in a multi-core or HPC setting (Fig.~\ref{fig:hic_flowchart}).

The automatic cut finder aims to find a cut strategy with minimal sampling overhead. Therefore, even for a cut budget $k_{max}$, the cut finder will return strategy for $k \ll k_{max}$ for a given device constraint, if it exists. While this minimizes sampling overhead, it can lead to highly unbalanced partitioning, i.e., given a circuit with $n$ qubits, if the cut strategy provides, say, two say subcircuits with $n_1$ and $n_2$ qubits where $n_1 \gg n_2$ (or vice versa), then the reduction of noise will be minimal since the larger subcircuit will end up acquiring most of the noise that was affecting the uncut circuit. In~\cite{basu2024fragqc} the authors show that balanced cutting, where $n_1 \simeq n_2$, often leads to better noise reduction. This observation motivates evaluating cutting strategies not only based on the number of cuts, but also on how effectively noise is distributed across subcircuits.

To avoid this pitfall, if $s_{min}$ and $s_{max}$ correspond to the smallest and largest connected component sizes (\# of qubits), we apply HIC (Algorithm~\ref{alg:cut_strategy}) for all device constraints $d \in [s_{min}, s_{max}]$, and select the one where the number of cuts $k \leq k_{max}$, and the weighted average layout score of all the subcircuits after placement is minimal. Selecting $d \in [s_{min}, s_{max}]$ can lead to subcircuit sizes that do not directly conform to the size of any connected component. However, since the size is bounded by \{$s_{min}, s_{max}$\}, it ensures that each subcircuit can be placed in at least one connected component. This guarantees that every evaluated cutting strategy remains compatible with the punctured hardware topology.

\begin{algorithm}[t]
\caption{HIC Method for the Selection of Cut Strategy}
\label{alg:cut_strategy}
\footnotesize
\begin{algorithmic}[1]
    \REQUIRE (i) set of connected components; (ii) pre-determined Z-scores $z_V$ and $z_E$ for qubits and 2-qubit connectivities; (iii) cut budget $k_{max}$, (iv) noise profile $\mathcal{N}$ of the hardware
    \ENSURE cut strategy with $k \leq k_{max}$ cuts where the difference in layout score of the subcircuits is minimal, or None if no such strategy exists

    \STATE $P_h \leftarrow$ Construct Punctured Coupling Map ($h$, $z_V$, $z_E$) for hardware $h$ via Algorithm~\ref{alg:punctured_coupling_map}
    \STATE $D \leftarrow$ set of all connected components in $P_h$
    \STATE cut\_strategies $\leftarrow$ []
    \FORALL{$d \in D$} 
        \STATE $c \leftarrow$ cut strategy obtained from automatic cut finder by setting the device constraint to be $d$
        \IF{number of cuts $k_c \leq k_{max}$}
            \STATE calculate $W_{s_c}$ via Algorithm~\ref{alg:calculate_ws}
            \STATE add $(d, c, W_{s_c})$ to cut\_strategies
        \ENDIF
    \ENDFOR
    
    \IF{|cut\_strategies| $== 0$}
        \RETURN None
    \ELSE
        \STATE $w \leftarrow \argmin_{c} \{W_{s_c}\}$
        \RETURN cut\_strategies[w]
    \ENDIF
\end{algorithmic}

\end{algorithm}

\subsection{Calculating Weighted Average Layout Score for Subcircuits}
\label{sec:weighted_avg}

The primary objective of circuit cutting in our framework is to reduce the effective noise experienced during execution by enabling subcircuits to be mapped onto higher-quality hardware layouts. As discussed before, \cite{basu2024fragqc} showed that equal partitioning usually leads to maximal lowering of noise. However, using $d \in [s_{min}, s_{max}]$ as the device constraint does not ensure equal partitioning. Consequently, selecting among multiple candidate cut strategies requires a quantitative criterion that reflects both the absolute quality of subcircuit placements and the balance of noise across them. So, we opt for the lowest overall noise when mapping the subcircuits to the components. 

We formalize this requirement through an objective function $W$ that evaluates a cut strategy based on both the quality of the individual subcircuit layouts and the uniformity of noise distribution across subcircuits. This function is defined as:

\begin{equation}
W \ = \ \alpha \frac{1}{n}\sum_{i=1}^S n_i s_i
\ + \
(1 - \alpha) \frac{1}{S}
\sum_{i=1}^S \left( \frac{1}{n}\sum_{j=1}^S n_j s_j - n_i s_i \right)^2
\end{equation}

where $S$ is the number of subcircuits, and $\alpha \in [0, 1]$.

Here, the first term computes a weighted average of the layout scores. We map each subcircuit $i$ to the best layout across all connected components. The quality of a layout is indicated by a layout score $s_i$ computed using the method of \cite{nation2023suppressing}, where lower values correspond to better layout quality. Since larger subcircuits tend to accumulate more noise, the score of each subcircuit is weighted by the number of qubits it contains. This term therefore captures the aggregate noise exposure across all subcircuits produced by a given cut strategy.

The second term addresses a limitation of using weighted averages alone -- highly unbalanced partitions may exhibit a deceptively low average score, despite concentrating noise in a single large subcircuit. Therefore, the second term computes the variance of the layout scores across subcircuits, penalizing strategies that result in uneven noise distribution. This ensures that no single subcircuit becomes a dominant noise bottleneck that degrades the accuracy of the reconstructed result. $\alpha$ parameterizes the trade-off between the best overall performance and a balanced noise distribution.

\begin{figure}[t]
    \centering
    \includegraphics[width=0.9\columnwidth]{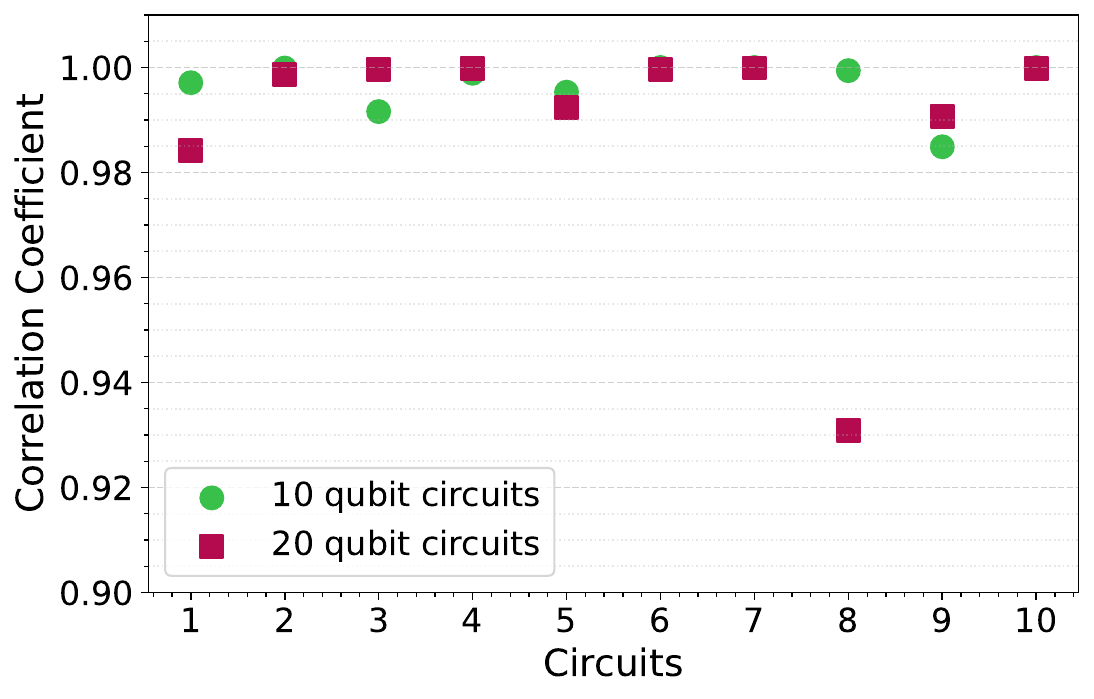}
    \caption{Correlation coefficient between norm-1 and norm-2 of the objective function $W$ for random 10-qubit (green circles) and 20-qubit (red squares) circuits. The high correlation across most instances justifies the simplification of the objective function to focus exclusively on norm-1 metric.}
    \label{fig:correlation_analysis}
\end{figure}

While the variance term is conceptually important for encouraging balanced partitioning, evaluating both terms increases computational complexity. To assess whether this complexity is necessary in practice, we numerically analyzed the correlation between the first (average) and second (variance) terms for multiple random circuits ranging from 10 to 50 qubits. For each of the case, the correlation coefficient \cite{schober2018correlation} always remains positive (infact $> 0.9$) which indicates that the norm-2 term varies monotonically with the norm-1 term. Fig.~\ref{fig:correlation_analysis} explicitly reports this coefficient for 20 such random circuits with 10 and 20 qubits. Consequently, our final objective function simplifies to the weighted average of the subcircuit layout scores, $W = \frac{1}{n}\sum_{i=1}^S n_i s_i$. This serves as a principled and more computationally efficient (and parallelly computable) criterion for comparing candidate cut strategies derived from different device constraints.

\begin{algorithm}[t]
\caption{Calculate Weighted Average Layout Score}
\footnotesize
\label{alg:calculate_ws}
\begin{algorithmic}[1]
    \REQUIRE (i) device constraint $d$; (ii) cut strategy $c$; (iii) noise profile of the hardware $\mathcal{N}$; (iv) set of connected components $D$
    \ENSURE weighted average score $W_{s_c}$

    \STATE $\{C_1, C_2, \dots, C_m\} \leftarrow \text{partition\_circuit}(c)$ 
    \COMMENT{Calls automatic cut finder to generate subcircuits}

    \STATE Initialize empty score list: $\text{scores} \leftarrow [\,]$

    \FOR{each subcircuit $C_i$}
        \FOR{each connected component graph $G_j \in D$}
            \STATE Compute layout score using the scoring function proposed in~\cite{nation2023suppressing}
            \STATE Append layout score to $\text{scores}$
        \ENDFOR
        \STATE $s_i \leftarrow \arg\min\limits_{j \in D} \text{scores}_{i,j}$ \COMMENT{Select best layout for $C_i$}
    \ENDFOR
    
    \STATE $W_{s_c} = \frac{1}{n} \sum_{i=1}^{S} n_i s_i$ \COMMENT{Compute weighted average over all subcircuits}

    \RETURN $W_{s_c}$
    
\end{algorithmic}
\end{algorithm}

Algorithm~\ref{alg:calculate_ws} details the procedure to compute $W$ for a given cut strategy and enables direct comparison across heterogeneous device constraints. For each device constraint $d$, the circuit is partitioned into subcircuits using the automatic cut finder. Each subcircuit is then mapped to its best possible layout across all connected components of the punctured coupling map, and the corresponding layout scores are aggregated using the weighted average metric. This objective function enables direct comparison of heterogeneous cut strategies and plays a central role in selecting the device constraint that best aligns circuit cutting with hardware noise non-uniformity.

\section{Experimental results}
\label{sec:result}

In this section, e present experimental results evaluating HIC on 20- and 50-qubit Clifford circuits, as well as application-level workloads from the Benchpress dataset \cite{nation2025benchmarking}. All results are obtained using noisy simulations on fake backends modeled after 133- and 156-qubit devices.
For all these cases, the observable is fixed to $\frac{1}{n} \sum_{i=0}^{n-1} Z_i$, where $n$ is the number of qubits. The random circuits were selected such that the ideal expectation value for the said observable does not go to zero, and to ensure the same, the QAOA circuit was mirrored. 
We compare HIC against equal partitioning using the automatic cut finder. 
This baseline reflects a noise-agnostic choice commonly used in practice when hardware-specific guidance is unavailable, making it an appropriate reference point for evaluating hardware-aware constraint selection.
This comparison allows us to isolate the impact of hardware-aware device-constraint selection on execution overhead and reconstructed result quality, i.e., whether modest reductions in the number of cuts can significantly lower overhead without severely degrading result quality. However, note that the number of cuts is not necessarily equal to the cut budget. A cut budget of $k_{max}$ implies that HIC will find a cut strategy with $k \leq k_{max}$ which leads to the minimum value of $W$. This is reflected in the results from Benchpress where the number of cuts obtained by HIC is often significantly lower than that of equal partitioning.

\subsection{Results for 20-qubit circuits}
\label{20_res}
We first evaluate HIC on a structured 20-qubit mirrored QAOA circuit, followed by unstructured random circuits. These experiments illustrate how hardware-aware device-constraint selection navigates the trade-off between execution cost and reconstruction accuracy more effectively than equal partitioning.

\subsubsection{Structured QAOA circuit (depth = 44)}
 
\begin{figure}[t]
    \centering
    \includegraphics[width=0.95\columnwidth]{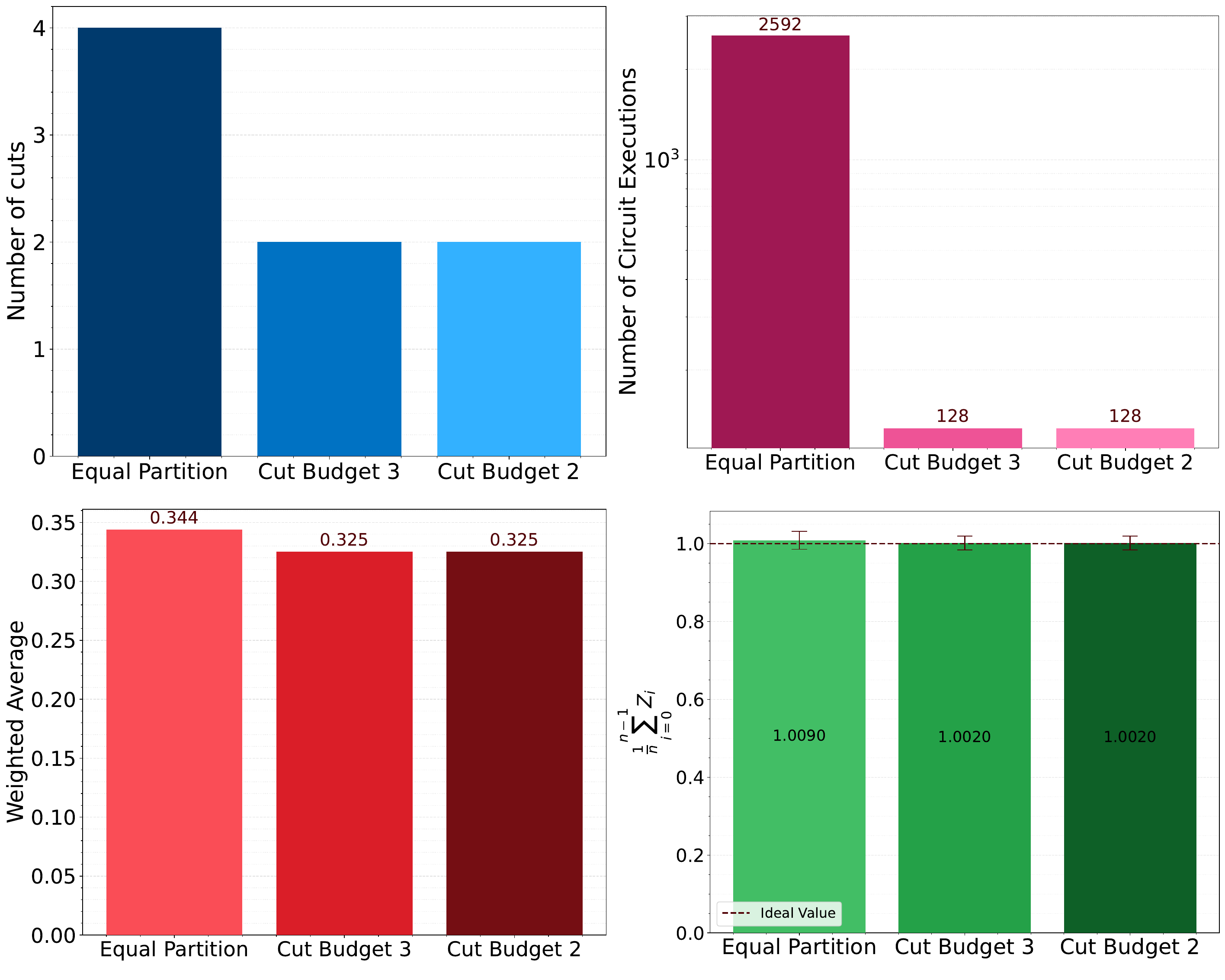} 
    \caption{Comparison of equal partitioning and different cut-budget strategies (HIC) for a structured 20-qubit mirrored QAOA circuit (depth = 44). The figures respectively show the number of cuts, number of circuit executions, evaluation of the weighted average, and the calculated expectation value of the observable $\frac{1}{n} \sum_{i=0}^{n-1} Z_i$. HIC with different cut budgets significantly reduce the overhead of circuit execution, while having a slight reduction in the weighted average score, and retaining the output quality.}
    \label{fig:qaoa_20q}
\end{figure}

Fig.~\ref{fig:qaoa_20q} demonstrates that HIC identifies a cut strategy with only 2 cuts compared to 4 under equal partitioning, reducing the number of circuit executions by approximately $20\times$ (from $2592$ to $128$). Despite this substantial reduction, HIC achieves a lower weighted layout score $W_s$, indicating better subcircuit placement, while retaining comparable (and marginally improved) output quality.
This behavior directly follows from the HIC framework, which systematically explores hardware-informed device constraints and selects the cut strategy minimizing the weighted layout score $W_s$, thereby jointly controlling sampling overhead and effective noise (Section~\ref{sec:our-proposed-method-of-cutting}).

While the structured QAOA circuit exhibits improvements in both execution overhead and reconstruction accuracy under our proposed method, such behavior cannot be expected to generalize uniformly across all workloads. In particular, for unstructured circuits, reducing the number of cuts introduces an inherent trade-off between execution cost and output quality. To illustrate this behavior, we next consider two random 20-qubit circuits: in the first case, a substantial reduction in execution overhead is achieved without materially affecting accuracy, whereas in the second, further reductions in overhead eventually lead to noticeable degradation in output quality.

\subsubsection{Random circuit 1 (depth = 128)}

\begin{figure}[t]
    \centering
    \includegraphics[width=0.95\columnwidth]{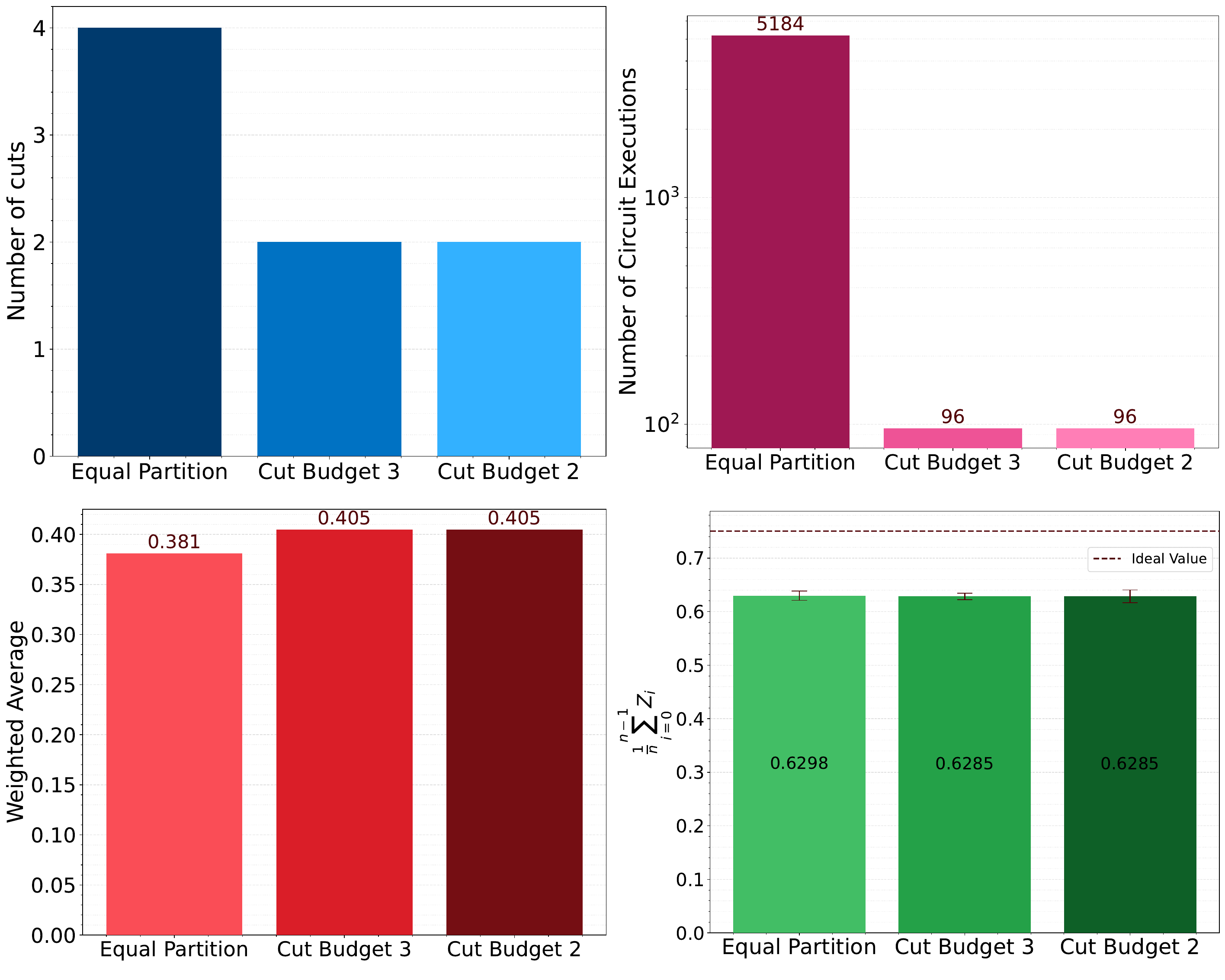} 
    \caption{Comparison of equal partitioning and different cut-budget strategies (HIC) for a 20-qubit random Clifford circuit (depth = 128). The figures respectively show the number of cuts, number of circuit executions, evaluation of the weighted average, and the calculated expectation value of the observable $\frac{1}{n} \sum_{i=0}^{n-1} Z_i$. HIC with different cut budgets significantly reduce the overhead of circuit execution, while having a slight increase in the weighted average score, and retaining the output quality.}
    \label{fig:rand_20q_seed_2}
\end{figure}

Fig.~\ref{fig:rand_20q_seed_2} compares equal partitioning and the proposed method for a 20-qubit random Clifford circuit. Under equal partitioning (device constraint = 10), the algorithm identifies 4 cuts, requiring $5184$ circuit executions and yielding a weighted average layout score $W_s = 0.381$. With a cut budget of $3$, the proposed method identifies a 2-cut strategy, reducing the number of circuit executions to $96$, a $\simeq 54\times$ reduction in quantum overhead. Although this results in a slightly higher weighted average ($W_s = 0.405$), the reconstructed expectation value remains comparable, indicating no material loss in output quality.

Reducing the cut budget to $2$ yields the same partition, while a cut budget of $1$ fails to produce a valid strategy. Consistent with the formulation in Section~\ref{sec:our-proposed-method-of-cutting}, the observed reconstruction accuracy closely follows the weighted average layout score, illustrating how $W_s$ captures the trade-off between execution overhead and result quality for hardware-aware device-constraint selection.

\subsubsection{Random circuit 2 (depth = 51)}
\begin{figure}[t]
    \centering
    \includegraphics[width=0.95\columnwidth]{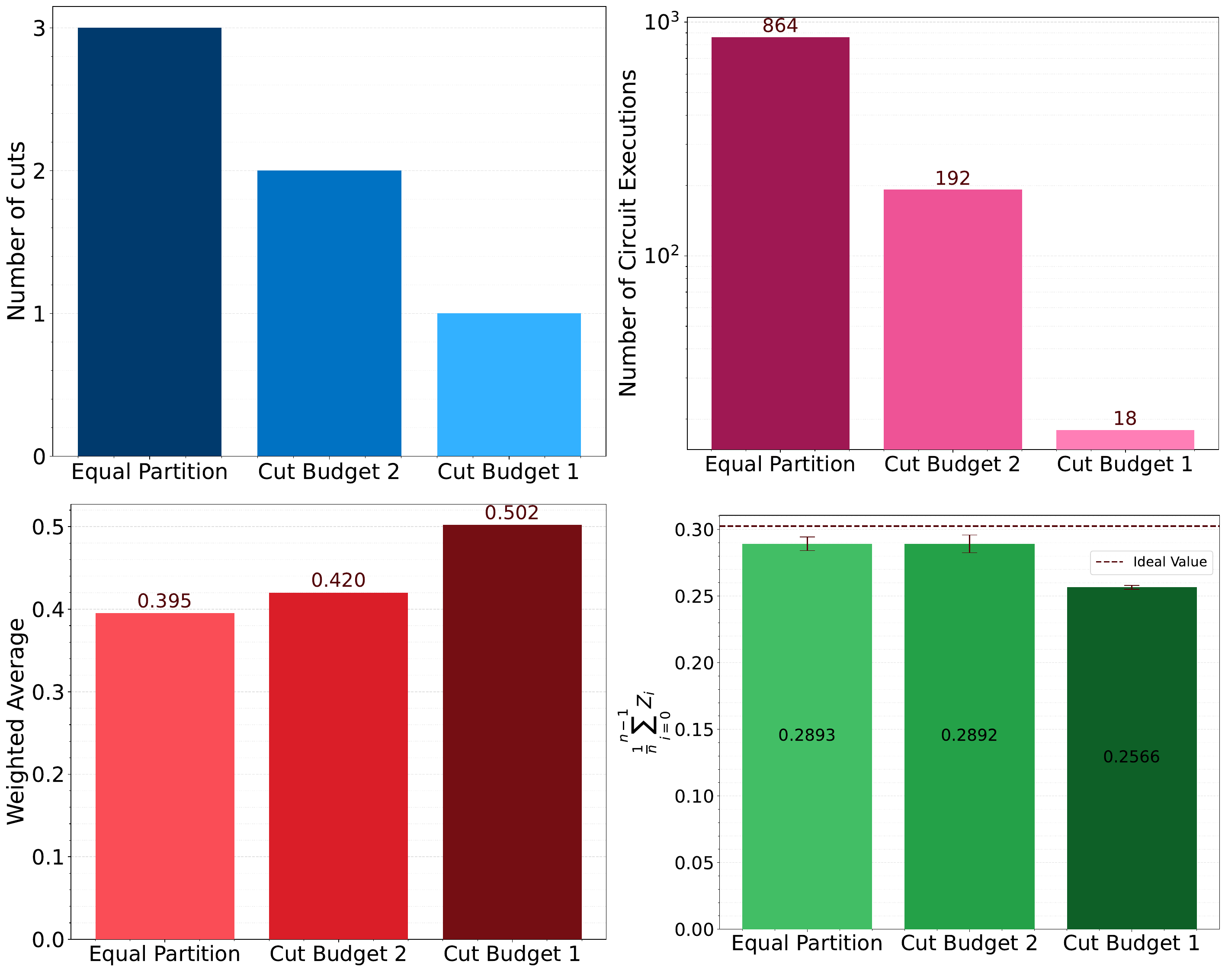} 
    \caption{Comparison of equal partitioning and different cut-budget strategies (HIC) for a 20-qubit random Clifford circuit (depth = 51). The figures respectively show the number of cuts, number of circuit executions, evaluation of the weighted average, and the calculated expectation value of the observable $\frac{1}{n} \sum_{i=0}^{n-1} Z_i$. HIC with cut budget $2$ significantly reduce the overhead of circuit execution, while having a slight increase in the weighted average score, and retaining the output quality. However, HIC with cut budget $1$ experiences a non-negligible increase in the weighted average score, and accordingly a degradation in output quality demonstrating that HIC cannot be used to arbitrarily reduce the execution overhead while retaining output quality.}
    \label{fig:rand_20q_seed_91}
\end{figure}

Fig.~\ref{fig:rand_20q_seed_91} presents results for a second 20-qubit random Clifford circuit. Under equal partitioning, the algorithm identifies $3$ cuts, requiring $864$ circuit executions with a weighted average layout score $W_s = 0.395$.
When the cut budget is reduced to $2$, the proposed method identifies a valid 2-cut strategy that lowers the number of circuit executions to $192$, a $\simeq 5\times$ reduction in the execution overhead, while increasing the weighted average layout score minimally to $W_s = 0.42$ and retaining comparable output quality.

Further reducing the cut budget to $1$ leads to a highly unbalanced partition. Although the number of circuit executions reduces to $18$ only, the weighted average layout score increases to $0.5$, and a significantly degradation in the output quality is observed. Taken together, this example highlights an important and expected property of the proposed method: HIC does not guarantee preservation of accuracy under aggressive cut-budget reduction. However, when a low-overhead, high-quality cut strategy exists within the feasible constraint space, HIC is able to identify it through hardware-aware device-constraint selection and optimization of the weighted layout score.

Collectively these results demonstrate that while circuit execution overhead cannot be reduced arbitrarily without impacting accuracy, the proposed HIC framework reliably identifies low-overhead cut strategies whenever such accuracy-preserving trade-offs exist within the feasible device-constraint space.

\subsection{50-qubit QAOA circuit result}
\label{50_qaoa_res}

\begin{figure}[h!]
    \centering
    \includegraphics[width=0.95\columnwidth]{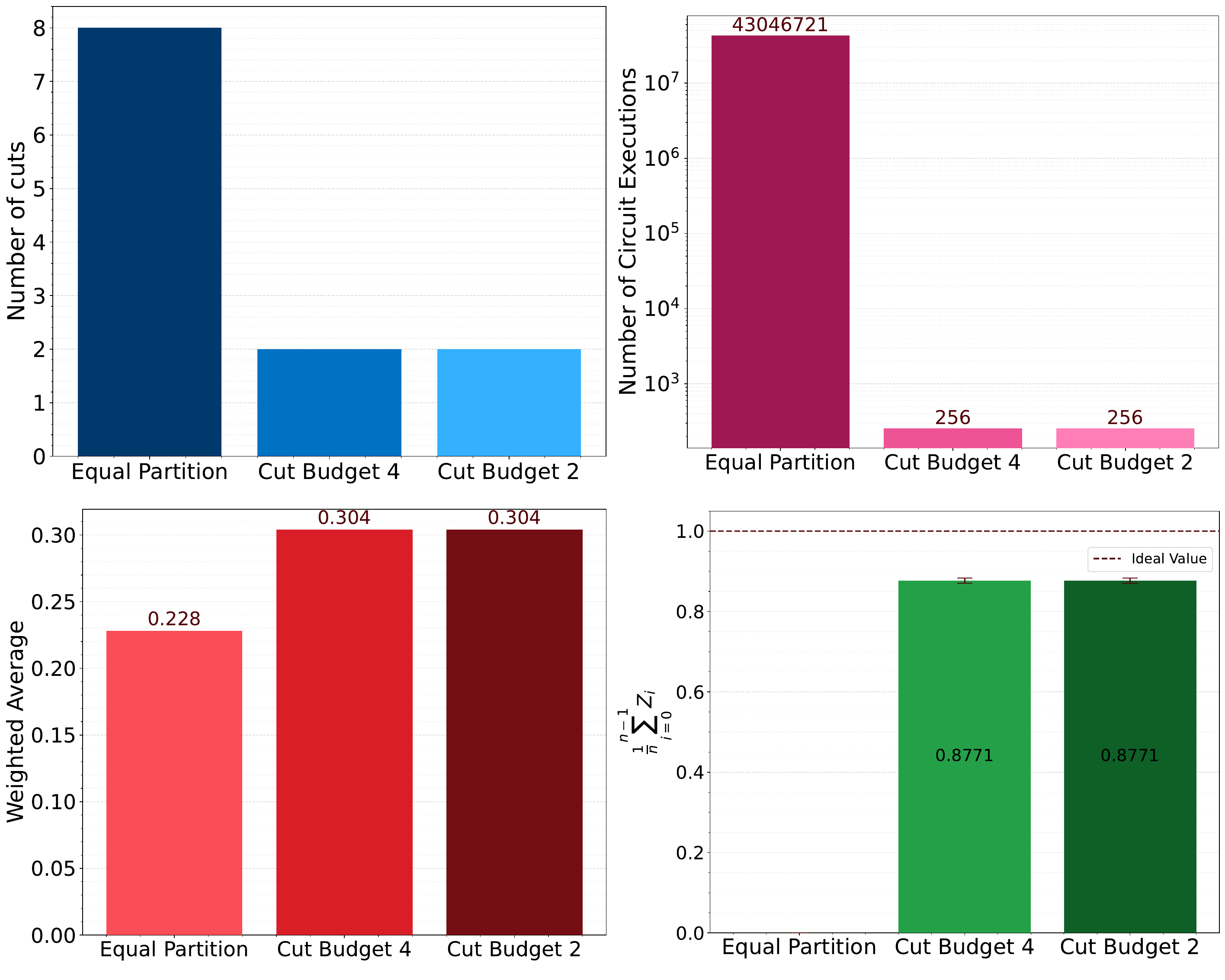} \caption{Comparison of equal partitioning and HIC strategies for a 50-qubit QAOA circuit (depth = 44). Equally partitioning the circuit leads to a substantial overhead in the number of circuit executions, making the reconstruction of the expectation value infeasible. With cut-budget 4, HIC identifies a 2-cut strategy. The number of circuit executions is significantly reduced, accompanied by an increase in the weighted average compared to the equal partitioning case. Reducing the cut budget to 2 resulted in the same strategy, as the cut budget of 4 itself suggested a 2-cut solution.}
    \label{fig:rand_50q}
\end{figure}

Fig.~\ref{fig:rand_50q} reports results for a 50-qubit mirrored QAOA circuit. Under equal partitioning, the automatic cut finder identifies 8 cuts, resulting in $43,046,721$ circuit executions. The weighted average layout score for equal partitioning is $W_s = 0.228$. 
In contrast, HIC identifies a valid 2-cut strategy, reducing the number of circuit executions to $256$ and bringing circuit cutting into a tractable region.
This reduction is accompanied by an increase in the weighted average to $W_s = 0.304$, causing an average deviation of $0.1229$ from the ideal expectation value. Since such a 2-cut strategy exists, the same result was observed when the cut budget was taken to be $4$ or $2$.

\begin{table*}[t]
\centering
\caption{Comparison of Equal Partitioning and HIC across various Benchpress circuits. The table reports the number of cuts, the total number of circuit executions, and the weighted average score ($W_s$) for both.  $W_{s_e} - W_{s_p}$ denotes the improvement in the weighted average using HIC, where negative values indicate a lower $W_s$ for the HIC strategy.  Results highlight that HIC achieves a significant reduction in number of circuit executions (up to 99.999\%) while maintaining or improving the weighted average.}

\begin{tabular}{l|r|r|r|r|r|r|r|r|r}
\hline
\textbf{Circuit} & \textbf{Qubits} 
& \multicolumn{3}{c|}{\textbf{Equal Partition}} 
& \multicolumn{3}{c|}{\textbf{HIC}} 
& \multirow{2}{*}{\bf $W_{s_e} - W_{s_p}$}
& \multirow{2}{*}{\bf \% Reduction in \# Circuits} \\
\cline{3-8}
& 
& \em  \# Cuts & \em  \# Circuits & $W_{s_{e}}$ 
& \em  \# Cuts & \em  \# Circuits & $W_{s_{p}}$ 
& & \\
\hline\hline

basis\_change & 3 
& 10 & 688,747,536 & 0.454
& 4 & 6,561 & 0.431
& 0.023 & 99.999\\
\hline

fredkin & 3 
& 8 & 43,046,721 & 0.379
& 4 & 6,561 & 0.287
& 0.092 & 99.985\\
\hline

linearsolver & 3 
& 4 & 6,561 & 0.306
& 2 & 81 & 0.264
& 0.042 & 98.77\\
\hline

qaoa & 3 
& 6 & 531,441 & 0.167
& 3 & 1,296 & 0.238
& -0.071 & 99.76\\
\hline

qft & 4 
& 8 & 43,046,721 & 0.266
& 4 & 20,736 & 0.274
& -0.008 & 99.95\\
\hline

seca & 11 
& 4 & 65,536 & 0.335
& 2 & 256 & 0.338
& -0.003 & 99.61\\
\hline

knn & 67 
& 2 & 256 & 0.556
& 1 & 16 & 0.575
& -0.019 & 93.75\\
\hline

\end{tabular}

\label{tab:bench_press_cut_comp}
\end{table*}

It is important to note that, for this 50-qubit case, we do not obtain a reconstructed expectation value for equal partitioning due to the prohibitive execution overhead associated with 8 cuts. The non-negligible increase in the weighted layout score ($W_s$) for HIC suggests some expected degradation in the output quality compared to equal partitioning. However, in practice, equal partitioning would require $43,046,721$ circuit executions, which is infeasible. On the other hand, HIC fundamentally enables circuit cutting to be executed at a reasonable and tractable overhead. In this sense, HIC improves over equal partitioning by making circuit cutting operationally feasible at this scale.

Another aspect is to verify the quality of result when the same resource is allowed to both equal partitioning and HIC for this case. In order to do so, we randomly sampled $256$ out of the $\simeq 43 \times 10^6$ circuits for equal partitioning, executed them, and reconstructed the expectation value. In accordance to the observations of \cite{majumdar2022error}, this yielded drastically worse result with an expectation value of $220.59 \pm 1.19$ over five trials. Comparing this to the ideal expectation value of $1$, and HIC achieving $0.87 \pm 0.006$ with the same number of circuit executions, show the superiority of HIC over equal partitioning approach for this case.

\subsection{Benchpress circuits result}
\label{bench_press_res}

To further assess the applicability of the HIC method beyond synthetic examples, we evaluate it on a set of application-level circuits from the Benchpress suite~\cite{nation2025benchmarking}. Many of these circuits require a large number of cuts under equal partitioning, resulting in an execution overhead that is prohibitively high. Consequently, for most Benchpress circuits, explicitly reconstructing expectation values under equal partitioning is infeasible. Given this limitation, we focus on two aspects: (i) the reduction in circuit execution overhead achieved by HIC relative to equal partitioning, and (ii) the corresponding change in the weighted average layout score $W_s$, which serves as a proxy for execution quality. As summarized in Table~\ref{tab:bench_press_cut_comp}, HIC consistently reduces the number of required circuit executions, often by more than two orders of magnitude, bringing the overhead into a tractable regime where circuit cutting can be practically executed.

We note that HIC sometimes leads to modest increase in the weighted layout score, but often at times it also reduces it. In Table~\ref{tab:layout_comparison} we explicitly show the number of qubits associated with the subcircuits, and the corresponding weighted layout score for the \emph{basis\_change} circuit, both for equal partitioning and HIC with a cut budget of 3. Therefore, for many circuits in Table~\ref{tab:bench_press_cut_comp}, HIC is expected to yield output quality comparable to, or better than, equal partitioning.
In scenarios where HIC leads to modest degradation in output quality, the loss remains limited relative to the substantial reduction in execution overhead.
On the other hand, the reduction in the overall circuit execution overhead for these circuits vary from $16\times$ to as high as $104976\times$.

Taken together, these results indicate that HIC does not eliminate the fundamental exponential overhead inherent to circuit cutting, but enables circuit cutting to be navigated in a regime where this execution overhead becomes feasible with minimal or no degradation in output quality.

\begin{table}[t]
\centering
\caption{The table compares the equal partitioning strategy (split into three subcircuits) with HIC (split into two subcircuits) on $basis\_change$ circuit for cut-budget 3. Individual scores represent the weighted layout scores for each subcircuit. HIC demonstrates a better output quality ($W_s = 0.431$) compared to Equal Partitioning ($W_s = 0.454$) as stated in Table~\ref{tab:bench_press_cut_comp}.}
\def\thickhline{\noalign{\hrule height1pt}} 

\begin{tabular}{c|c|c|c}
\hline
\textbf{Method} & \textbf{Subcircuit} & \textbf{Selected Layout} & \textbf{Score} \\
\hline\hline
\multirow{3}{*}{\em Equal Partition}
& 0 & [0] & 0.4221 \\ \cline{2-4}
& 1 & [0] & 0.5186 \\ \cline{2-4}
& 2 & [0] & 0.4221 \\ \cline{2-4}
& \multicolumn{2}{c|}{$W_{s}$} & 0.4542 \\
\thickhline
\multirow{2}{*}{\em Cut Budget = 3}
& 0 & [0] & 0.4217 \\ \cline{2-4}
& 1 & [0,1] & 0.4353 \\ \cline{2-4}
& \multicolumn{2}{c|}{$W_{s}$} & 0.4308 \\
\hline
\end{tabular}
\label{tab:layout_comparison}
\end{table}

Having established that HIC enables circuit cutting to be executed at practical overheads across a broad range of workloads, we now compare its performance against existing circuit cutting approaches to quantify the overhead of quantum circuit execution, as well as the classical pre-processing and post-processing.

\section{Related Works}
\label{sec:related}

Having established the HIC methodology for hardware-aware device-constraint selection, we now position our contribution relative to prior work on circuit cutting and cut-strategy selection. In this section, we contrast our proposed HIC method against existing circuit cutting strategies. The first proposal of circuit cutting~\cite{peng2020simulating} did not consider the problem of identifying cut locations; cuts were derived manually for the circuits under study. Subsequently, Tang \emph{et al.}~\cite{tang2021cutqc} formulated cut-location selection as an integer linear program, with relaxation techniques, to minimize classical reconstruction overhead. Perlin \emph{et al.}~\cite{perlin2021quantum} investigated maximum likelihood tomography to improve reconstruction quality in circuit cutting, at the cost of increased classical overhead, and Majumdar \emph{et al.}~\cite{majumdar2022error} further combined such reconstruction techniques with error mitigation to improve output fidelity under noise.

However, all of the above approaches consider \emph{wire cuts} exclusively, where circuit partitioning is achieved by severing qubit wires. Mitarai \emph{et al.}~\cite{mitarai2021constructing, mitarai2021overhead} introduced \emph{gate cutting}, where long-range multi-qubit gates are decomposed into sequences of single-qubit operations and measurements. The Automatic Cut Finder~\cite{qiskit-addon-cutting} unifies these two paradigms by performing a heuristic tree search over both wire and gate cuts to minimize sampling overhead, given a user-specified device constraint. Our proposed HIC uses the Automatic Cut Finder explicitly as a subroutine, allowing it to access its advantages while focusing on systematic device-constraint selection rather than cut-location optimization.

Several works have explored algorithmic cut selection. Basu \emph{et al.}~\cite{basu2022qer} proposed a learning-based approach to predict cut locations that minimize subcircuit noise, limited to bipartitioning, requiring retraining as hardware noise drifts, with a training bottleneck towards scalability. FragQC~\cite{basu2024fragqc} extends this idea by enforcing balanced min-cut partitioning to equalize noise across subcircuits; however, it assumes a uniform hardware noise model, an assumption that does not hold for contemporary devices (Fig.~\ref{fig:noise_profile}). In contrast, HIC explicitly targets the non-uniformity of hardware noise by deriving device constraints from punctured coupling maps rather than imposing balance constraints or learned heuristics.

Other approaches address orthogonal aspects of circuit cutting. Chen \emph{et al.}~\cite{chen2023online} propose identifying \emph{golden cut points} to reduce the number of measurement bases required for reconstruction. This technique is circuit specific and can be applied equally to subcircuits produced by HIC, making it complementary rather than competing. Additionally, recent theoretical work explores circuit cutting with classical communication or joint partitioning strategies~\cite{piveteau2023circuit, brenner2025optimal, schmitt2025cutting}. However, current quantum hardware does not yet support classical communication within gate execution, and hence these methods are not immediately implementable.

As a result, in the following section we restrict quantitative comparison to CutQC~\cite{tang2021cutqc} and FragQC~\cite{basu2024fragqc}, as these represent fully implemented, widely used cut-finding strategies that directly overlap with HIC in scope and operate under comparable hardware and execution assumptions.

\section{Comparing the performance of HIC with existing cut finding strategies}
\label{sec:cutqc_vs_hic}

\begin{table}[t]
\centering
\setlength{\tabcolsep}{1.5pt} 
\def\thickhline{\noalign{\hrule height1pt}} 

\caption{Comparison of Pre-processing and Post-processing time, and number and type of cuts for circuit cutting using CutQC, FragQC, and HIC on 20-qubit circuits}

\begin{tabular}{C{1.75cm}|l|l|R{1.3cm}|R{1.3cm}|r|r}
\hline
\textbf{Circuit} & \multicolumn{2}{c|}{\textbf{Method}} & \textbf{Pre-proc. Time (s)} & \textbf{Post-proc. Time (s)} & \multicolumn{2}{c}{\textbf{Number of Cuts}} \\
\cline{6-7}
 & \multicolumn{2}{c|}{} & & & \textbf{Wire-cut} & \textbf{Gate-cut} \\
\hline\hline

\multirow{4}{*}{\bf \makecell{20-qubit\\QAOA}}
& \multicolumn{2}{l|}{\em CutQC} & 0.388 & 3.223 & 2 & - \\ \cline{2-7}
& \multicolumn{2}{l|}{\em FragQC} & 3.393 & - & 38 & - \\ \cline{2-7}
& \multirow{2}{*}{\em HIC} & Serial & 31.695 & \multirow{2}{*}{1.264} & \multirow{2}{*}{2} & \multirow{2}{*}{-} \\ \cline{3-4}
& & Parallel & 3.60 & & & \\ \thickhline

\multirow{4}{*}{\bf \makecell{20-qubit\\ Random\\ Circuit 1}}
& \multicolumn{2}{l|}{\em CutQC} & 0.537 & 3.894 & 2 & - \\ \cline{2-7}
& \multicolumn{2}{l|}{\em FragQC} & 5.412 & - & 80 & - \\ \cline{2-7}
& \multirow{2}{*}{\em HIC} & Serial & 19.423 & \multirow{2}{*}{1.007} & \multirow{2}{*}{1} & \multirow{2}{*}{1} \\ \cline{3-4}
& & Parallel & 5.83 & & & \\ \thickhline

\multirow{4}{*}{\bf \makecell{20-qubit\\ Random\\ Circuit 2}}
& \multicolumn{2}{l|}{\em CutQC} & - & - & - & - \\ \cline{2-7}
& \multicolumn{2}{l|}{\em FragQC} & - & - & - & - \\ \cline{2-7}
& \multirow{2}{*}{\em HIC} & Serial & 3.316 & \multirow{2}{*}{0.259} & \multirow{2}{*}{-} & \multirow{2}{*}{1} \\ \cline{3-4}
& & Parallel & 1.77 & & & \\ \hline

\end{tabular}
\label{tab:cutting_comparison}
\end{table}

Some of the widely used approaches for circuit cutting include CutQC~\cite{tang2021cutqc}, FragQC~\cite{basu2024fragqc}, and the Automatic Cut Finder~\cite{qiskit-addon-cutting}. The goal of this section is to quantify the practical impact of HIC relative to these approaches under identical circuit and noise settings. Note that the weighted layout score $W_s$ is used as a hardware-aware proxy for noise exposure during cut-strategy selection, not as a direct measure of output correctness. All reconstructed expectation values are computed using identical shot budgets across methods to ensure fair comparison of execution overhead and result quality.

The Automatic Cut Finder requires the user to specify the maximum number of qubits per subcircuit (the \emph{device constraint}), which is generally nontrivial to determine \emph{a priori}.
An ill-defined device constraint can lead to significant increase in the required number of circuit executions (see Table~\ref{tab:cut_stats}). 
Because the Automatic Cut Finder provides no algorithmic mechanism for selecting this parameter, its effectiveness depends heavily on user intuition and prior knowledge of both the circuit and the hardware.
In this work, the Automatic Cut Finder serves as an underlying primitive within HIC rather than as a standalone baseline.
HIC is explicitly designed to automate and systematize device-constraint selection under realistic, non-uniform hardware noise profiles. 
Consequently, a direct comparison with the Automatic Cut Finder would primarily reflect differences in manual parameter tuning rather than intrinsic cutting capability.
We therefore focus our comparison on CutQC and FragQC, which represent established approaches that rely on user-specified or heuristic constraints, and evaluate how HIC impacts both the feasibility and efficiency of the resulting cut strategies

We now compare HIC against CutQC~\cite{tang2021cutqc} (using the implementation from \texttt{qiskit-addon-cutting} version 0.10.0) and FragQC~\cite{basu2024fragqc}. 
The comparison focuses on classical pre-processing time (cut discovery), post-processing time (expectation-value reconstruction), and the number and type of cuts produced.
In particular, we highlight the trade-off between classical pre-processing cost and the ability to reduce quantum execution overhead through a richer space of admissible cut strategies. All methods are evaluated on the same circuits shown in Section~\ref{sec:result}, allowing direct comparison under consistent experimental conditions.

First, note that both CutQC and FragQC require the user to provide a \emph{maximum subcircuit width}, which is equivalent to the device constraint required by Automatic Cut Finder and, in general, is not known \emph{a priori}. This already highlights a fundamental limitation shared by both CutQC and FragQC: in the absence of an algorithmic mechanism for selecting device constraints, their effectiveness depends heavily on user intuition and manual tuning. 
In this comparison, the maximum subcircuit width for CutQC was derived using HIC and refined through limited trial runs around that value.
This refinement was necessary because CutQC supports only wire cuts, whereas HIC may produce strategies involving both wire and gate cuts. FragQC, which recursively applies balanced partitioning on top of CutQC, inherits these same constraints.

Table~\ref{tab:cutting_comparison} summarizes the classical pre-processing time, post-processing time, and the number and type of cuts across all evaluated methods and circuits. While HIC exhibits higher serial pre-processing time compared to CutQC and FragQC, this overhead arises from its systematic exploration of hardware-informed device constraints rather than a single heuristic search. Importantly, when executed in parallel, HIC's pre-processing time becomes comparable to FragQC, while retaining its broader cut expressivity. For this paper, the parallel execution was performed on a system equipped with an Apple M4 chip, 16~GB RAM, and 10 CPU cores.

\begin{figure*}
    \centering
    \includegraphics[width=0.95\textwidth]{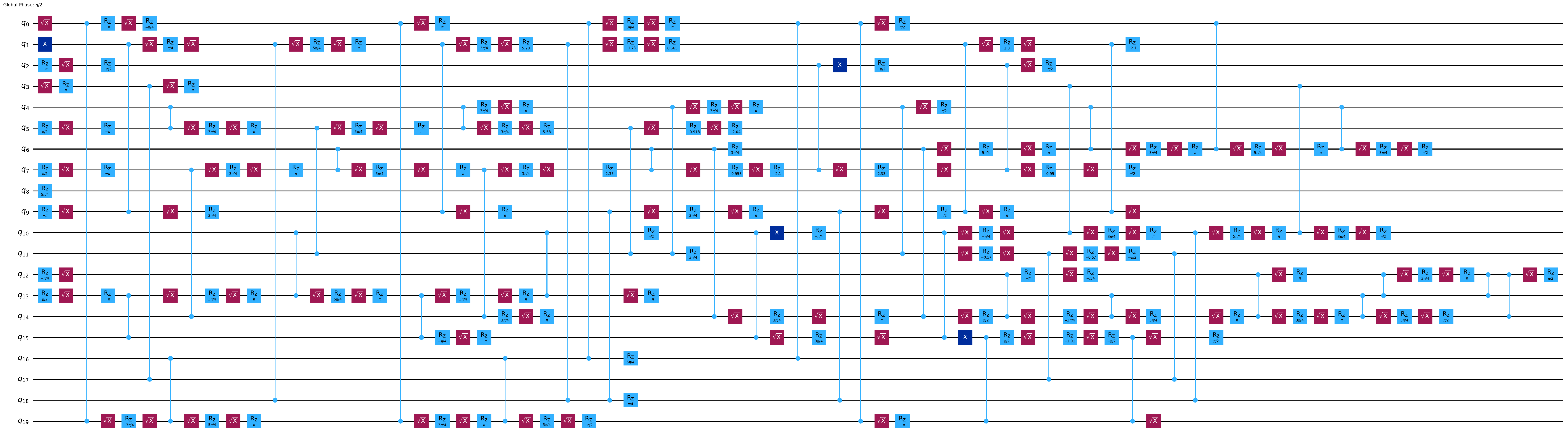} \caption{A 20-qubit random circuit with a 2-qubit gate from the first qubit to the last qubit. Therefore, any cut strategy has to partition this gate to create disjoint subcircuits -- thus explaining why wire-cut-only strategies fail.}
    \label{fig:rand_20q_seed91_cutqc}
\end{figure*}

For the structured 20-qubit QAOA circuit, the pre-processing times of CutQC and FragQC are lower than HIC's serial execution. However, as stated before, when HIC explores the different subcircuit sizes in parallel, its execution time becomes comparable to FragQC. Nevertheless, FragQC produces a prohibitively large number of cuts (38), rendering circuit execution infeasible, while CutQC requires manual tuning to identify a wire-cut strategy with 2 cuts. In contrast, HIC automatically identifies a 2-cut strategy without manual intervention, and directly supports expectation-value reconstruction rather than full probability reconstruction. As a result, HIC's post-processing cost is $2.55\times$ lower than that of CutQC for this circuit. A comparison with FragQC was not feasible due to the impractical number of circuit executions it would take $(\simeq 16^{38}$) before reconstruction.

For the first 20-qubit random circuit, both CutQC and FragQC fail to produce practically useful cut strategies. FragQC requires 80 cuts, while CutQC identifies a trivial partition where one subcircuit contains the entire circuit and the other is a trivial single qubit circuit with no gates, effectively nullifying the benefit of cutting. In contrast, HIC identifies a balanced partition using only 1 gate cut and 1 wire cut, significantly reducing the sampling overhead. 
This demonstrates that the additional classical overhead incurred by HIC enables access to a richer and more effective cut-strategy space that is fundamentally inaccessible to wire-cut-only methods.

For the second 20-qubit random circuit (see Fig.~\ref{fig:rand_20q_seed_91}), neither CutQC nor FragQC can identify any valid cut strategy, as the circuit topology necessitates a gate cut. This can be easily validated: notice that the first 2-qubit gate is between the first and last qubits, and therefore, any cut strategy needs to partition this gate, thus requiring gate cutting. HIC, by explicitly allowing gate cuts, successfully identifies a feasible strategy with a single gate cut, making circuit cutting possible for a workload where existing methods fail entirely.

Overall, these results indicate that while HIC may incur higher classical pre-processing cost in its serial form, it (i) consistently produces feasible and lower-overhead cut strategies in cases where CutQC and FragQC either fail or incur impractical sampling overheads, and (ii) achieves classical pre-processing times comparable to FragQC when the search is parallelized.

Taken together, the results in this and the previous section establish HIC as a practical and enabling approach to circuit cutting -- one that transforms cutting from a technique that is theoretically appealing, or applicable only to small-scale toy circuits, into an operationally feasible tool on realistic, noisy quantum hardware.

\section{Discussion and Conclusion}
\label{sec:conclusion}

In this article, we introduced Hardware-Inspired Cutting (HIC), a circuit cutting framework that systematically reduces quantum execution overhead by exploiting hardware noise non-uniformity. HIC focuses on identifying cut strategies that balance sampling overhead and effective noise, as quantified by the weighted layout score $W_s$. In this work, we prioritize result quality as the primary objective of circuit cutting, rather than minimizing execution latency or maximizing parallelism. As a result, HIC may favor cut strategies that place multiple subcircuits on the same connected component if doing so yields higher-quality outcomes, even when additional parallel execution opportunities exist. While subcircuits mapped to distinct connected components could, in principle, be executed in parallel, either within a single device or across multiple devices, we do not consider such parallelism in this study. Extending the objective function to explicitly incorporate execution latency, queuing delays, or inter- and intra-device scheduling remains an interesting direction for future work. Importantly, the current formulation of HIC is naturally compatible with such extensions, as the evaluation of candidate device constraints and cut strategies is already fully parallelizable in the classical pre-processing stage.

A key aspect of this study is that all observed improvements arise solely from leveraging hardware noise non-uniformity during device-constraint selection. We do not combine HIC with complementary techniques such as Operator Backpropagation, which is known to reduce cut overhead~\cite{pal2025low}. While such technique could further lower execution cost, we deliberately isolate the contribution of noise-aware device-constraint selection in this paper to clearly demonstrate its standalone impact. Another promising direction for future work is to learn a predictive relationship between changes in the weighted layout score $W_s$ and the resulting degradation in output quality, for example using lightweight machine learning models. Such a model could enable users to estimate, in advance, whether a given increase in $W_s$ relative to equal partitioning is likely to yield acceptable accuracy, thereby guiding expectations about when HIC is likely to produce high-quality results in practice.

In summary, this work demonstrates that hardware-aware device-constraint selection can fundamentally impact the practical viability of circuit cutting. By systematically exploiting noise non-uniformity, HIC enables circuit cutting to be executed at realistic overheads, with this paper demonstrating up to 50-qubit circuits, including  scenarios where conventional strategies become impractical. By making circuit cutting operationally feasible on realistic noisy hardware, this work brings circuit cutting closer to sustained, practical use in near-term quantum computing.

\textbf{Code availability}: The codes used for the experiments reported in this manuscript are available at 
\url{https://github.com/dream-lab/quantum-hic/}.

\IEEEtriggeratref{12}
\bibliographystyle{ieeetr}
\bibliography{IEEEabrv}

\end{document}